\documentclass[prl,reprint,nofootinbib]{revtex4-1}
\usepackage{graphicx} % Remove [demo] to get actual pictures
\usepackage{amsmath}

\usepackage{color}

\newcommand{\be}{\begin{equation}}
\newcommand{\ee}{\end{equation}}

\newcommand{\tr}[1]{\text{Tr}\big[{#1}\big]}
\begin{document}
	
	\title{Quantum Hydrodynamics in Spin Chains with Phase Space Methods}
	%\author{}
	
	\author{Jonathan Wurtz}
	\email[Corresponding author: ] {jwurtz@bu.edu}
	\affiliation{Department of Physics, Boston University, 590 Commonwealth Ave., Boston, MA 02215, USA}

	\author{Anatoli Polkovnikov}
	\affiliation{Department of Physics, Boston University, 590 Commonwealth Ave., Boston, MA 02215, USA}
	%\date{Started 6/27/2018}
	\date{8/27/2018}

	\begin{abstract}
		Connecting short time microscopic dynamics with long time hydrodynamics in strongly correlated quantum systems is one of the outstanding questions. In particular, it is very difficult to determine various hydrodynamic coefficients like the diffusion constant or viscosity starting from a microscopic model: exact quantum simulations are limited to either small system sizes or to short times, which are insufficient to reach asymptotic behavior. In this Letter, we show that these difficulties, at least for a particular model, can be circumvented by using the cluster truncated Wigner approximation (CTWA), which maps quantum Hamiltonian dynamics into classical Hamiltonian dynamics in auxiliary high-dimensional phase space. We apply CTWA to a XXZ next-nearest-neighbor spin 1/2 chain and find behavior consisting of short time spin relaxation which gradually crosses over to emergent diffusive behavior at long times. For a random initial state we show that CTWA correctly reproduces the whole spin spectral function. Necessary in this construction is sampling from properly fluctuating initial conditions: the Dirac mean-field (variational) ansatz, which neglects such fluctuations, leads to incorrect predictions.
		
	\end{abstract}

	\maketitle

	\textit{Introduction}--Thermalization of quantum systems has recently become a focus of active research both theoretical and experimental \cite{Kaufman794,Neill2016,Gogolin2016,BORGONOVI20161}. It has been realized that quantum chaos and emerging relaxation to equilibrium is encoded in the structure of many-body eigenstates of generic quantum Hamiltonians \cite{Santos2010,polkovnikov2011_rev,DAlessio2015}. Despite this progress most theoretical studies of quantum thermalization is either confined to small systems amenable to exact diagonalization \cite{Mukerjee2011,Steinigeweg2014,Agarwal2015,Richter2018} or to more phenomenological hydrodynamic and kinetic approaches \cite{Moeckel2008,Lux2014,Ljubotina2017}. Recently new approaches like a novel Gaussian variational approach for quantum impurity systems \cite{Ashida2018}, the time-dependent variational ansatz (TDVP) \cite{Gobert2005,Kramer2008,Leviatan2017,Kloss2018}, and the cluster truncated Wigner approximation (CTWA) \cite{Wurtz2018} were proposed as viable tools for studying long time relaxation of quantum systems to thermal equilibrium. The latter two approaches share a common feature that they approximate long time quantum dynamics with effective non-linear classical dynamics in a high-dimensional phase space, which can be systematically increased to ensure convergence of the results to the correct ones. This mapping reduces complexity of simulations of quantum dynamics from exponential to polynomial in the system size, which should be intuitively sufficient for proper description of long-time large scale hydrodynamic behavior.  One key feature of the CTWA approach is that unlike mean field approaches it contains fluctuating initial conditions distributed according to the appropriate (Wigner) function describing the initial state. Therefore the information about observables and correlations in CTWA can be only obtained through averaging over many trajectories, each describing a different effective mean field evolution.

	Microscopically, hydrodynamic coefficients can be expressed through appropriate non-equal time correlation functions. In equilibrium there are various thermodynamic relations between transport and response coefficients such as fluctuation-dissipation relation \cite{landau2013statistical}, drift-diffusion Einstein and Onsager relations \cite{Onsager1931} and others. These thermodynamic identities imply that a proper formalism describing thermalization should not only explain relaxation of various observables to their thermal values but also proper asymptotic behavior of non-equal time correlation functions and the dynamic structure factor $S(k,\omega)$. 
	
	In this work, using phase space methods developed previously \cite{Wurtz2018}, we study $S(k,\omega)$
	for a generic spin $1/2$ next-nearest-neighbor XXZ chain at infinite temperature. In particular, we correctly recover both its high and low frequency asymptotics: low frequencies corresponding to hydrodynamic diffusive relaxation, while high frequencies describe short time coherent quantum excitations. These methods also predict a nontrivial crossover between the two asymptotic regimes. While high frequency behavior can be obtained using exact diagonalization in relatively small systems, the correct description of low frequencies requires access to system sizes which are beyond the range of existing methods. We also show that noise in initial conditions is crucial for correctly predicting the structure factor and the spin diffusion constant and that the cluster mean field dynamics, which can be obtained from CTWA by suppressing noise, leads to incorrect predictions.

	The cluster truncated Wigner approximation (CTWA), recently introduced in Ref.~\cite{Wurtz2018}, is the specific phase space method used in this Letter. The CTWA amounts to first splitting a system of interest into disconnected clusters of spins (labeled by $``i"$) and interpreting the complete set of Hermitian operators inside each cluster $G=\{\hat X^{i}_\alpha\}$ as classical phase space variables  $x^{i}_\alpha$ (see Supplementary Information for details). Quantum operators, including the observables and the Hamiltonian, are mapped to functions of these variables. To describe the dynamics of the system, an ensemble of points is independently evolved in time according to a non-linear classical Hamiltonian induced from the quantum Hamiltonian, with the initial conditions drawn from a Gaussian probability distribution reproducing averages and fluctuations of the operators $\hat X^i_\alpha$ in the initial state. To compute time dependent expectation values of observables at time $t$ (or similarly the non-equal time correlation functions) we average the corresponding functions evaluated on this ensemble of classical trajectories. This sampling of initial conditions is critical: because the dynamics are nonlinear, nearby trajectories generically diverge in time in a way which encodes spreading of correlations and entanglement in phase space. If a single trajectory is used, i.e. all fluctuations are suppressed to zero, then the CTWA reduces to the cluster Dirac mean-field (variational) approximation. %\jw{\textbf{[REFS]}}
	
	% JW 8/9- Way too much time was spent making this sound right =/
	The CTWA is approximate, but improves as the cluster size, and thus the dimensionality of the phase space, increases. This is because the method treats the dynamics within a cluster exactly: it captures all entanglement and correlations within the expanded phase space. Dynamics between clusters are nonlinear and do not capture any quantum correlations per point in phase space: in this way time evolution of individual points are mean field projective dynamics. This means that in the limit of the cluster size as the system size the method recovers the exact result, with the caveat that the phase space dimensionality is now exponential in the system size. The computational difficulty scales as $2^{L+1}$ in the cluster size, eg evolution of wave functions.

	One of the quantities that the CTWA can approximately reproduce is the infinite temperature symmetric non-equal time correlation function of two spin operators $\hat A$ and $\hat B$:
	\begin{multline}
	G_{AB}(t,t')={1\over \mathcal{D}}{\rm Tr}[\hat A(t)\hat B(t')]={1\over 2\mathcal{D}}{\rm Tr}[\{\hat A(t),\hat B(t')\}_+],\\ 
	\equiv {1\over 2\mathcal{D}}\sum_n\langle \psi_n|\hat A(t)\hat B(t')+\hat B(t') \hat A(t)|\psi_n\rangle,
	\end{multline}
	where $\mathcal D$ is the total Hilbert space dimension and $\{|\psi_n\rangle\}$ is a complete basis of states. For computational purposes we sample the correlation function over spin states randomly polarized along the Z-axis. Under the CTWA the expectation value appearing in the equation above for each of the states $|\psi_n\rangle$ is approximately reproduced as:
	
	\begin{equation}\label{eq:symm_corr}
	\langle\psi_n | \{\hat A(t),\hat B(t')\}_+|\psi_n\rangle \approx 2\overline{A(\vec x(t))B(\vec x(t'))},
	\end{equation}
	where $\vec x(t)$ denote coordinates of a specific phase space point evolved to time $t$ drawn from the initial probability distribution, and $A(\vec x(t))$ is the Weyl symbol of the operator $\hat A$ evaluated at $\vec x(t)$. The overline denotes averaging with respect to Gaussian initial conditions at $t=0$ corresponding to the state $|\psi_n\rangle$ (see Supplementary Information and Ref.~\cite{Wurtz2018} for details).

	\begin{figure}
		\includegraphics[width=\linewidth]{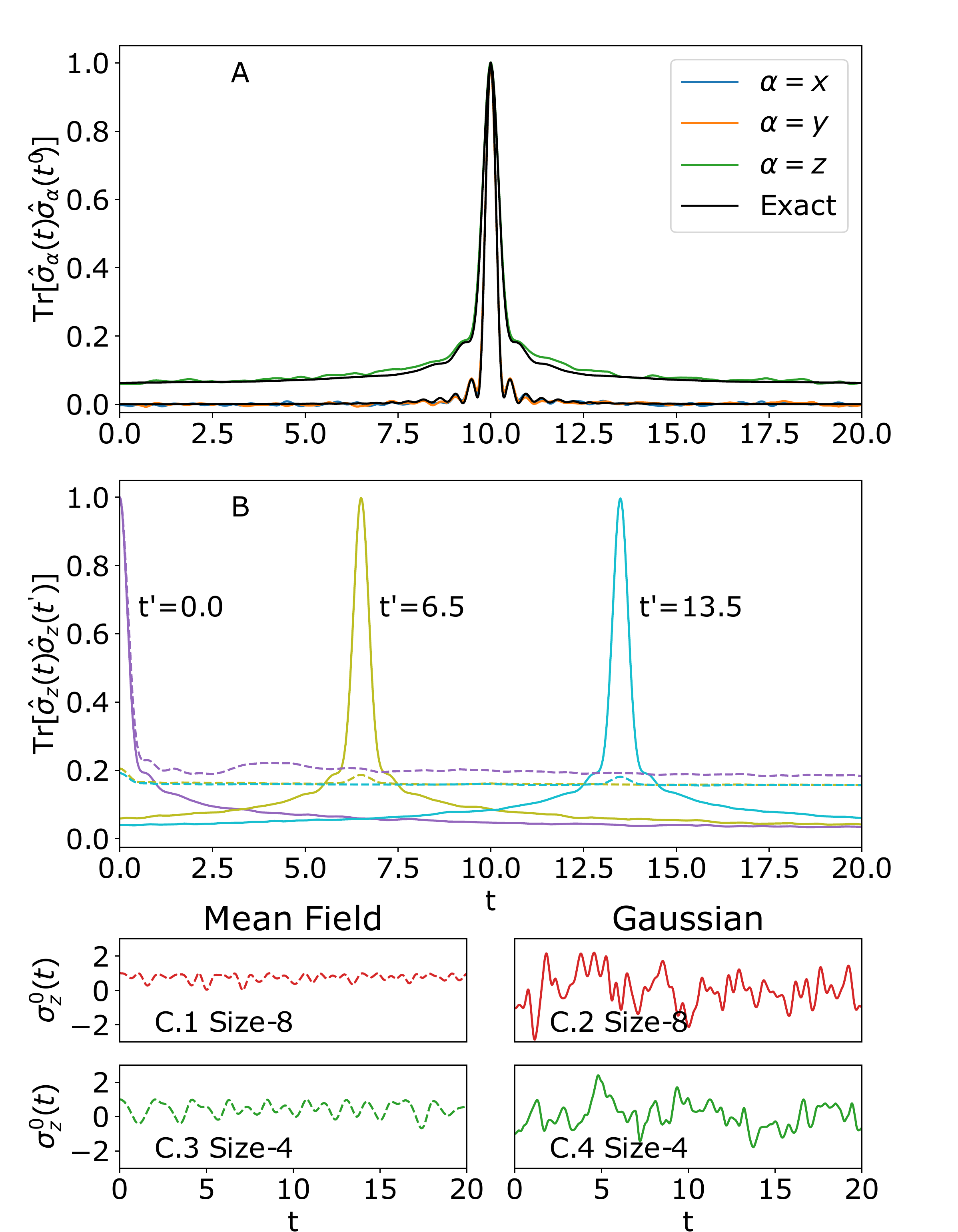}
		\caption{\textbf{Non-equal time spin-spin correlation functions of the next-nearest neighbor XXZ chain of eq. \ref{eq:nnn_XXZ} at infinite temperature }. \textbf{(A)} Correlation function for $t'=10$, compared to exact results. $(L,N)=(8,16)$ \textbf{(B)} Correlation function for $(L,N)=(8,64)$, which shows that the correlation function is well captured for offsets $t'$. Mean field (dashed) does not capture correctly, emphasizing importance of fluctuations.
			\textbf{(C)} Time traces of individual points in phase space for a typical (dashed) mean-field and (solid) Gaussian initial condition. Fluctuations persist at all times for Gaussian, but are exponentially small in the cluster size for the mean-field case. Cluster and system sizes are $(L,N)=(8,64)$.} \label{fig:offset_timecorr}
		% JW - These were the least-ugly colors I could find without going really deep into color schemes. Also, it sort've bothers me that "Size-8" slightly overlaps with the time trace, but I can't figure out how to overcome this.
	\end{figure}

	To demonstrate how the method works we choose a particular next-nearest neighbor spin-1/2 XXZ model with periodic boundary conditions, which conserves the total Z magnetization but has no extensive symmetries.
	\begin{eqnarray}\label{eq:nnn_XXZ}
	\hat H = \sum_{i}^N \hat \sigma^i_{x}\hat \sigma^{i+1}_x+\hat \sigma^i_{y}\hat \sigma^{i+1}_y+\Delta \hat \sigma^i_{z}\hat \sigma^{i+1}_z&&\\
	+\gamma  \sum_{i}^N \hat \sigma^i_{x}\hat \sigma^{i+2}_x+\hat \sigma^i_{y}\hat \sigma^{i+2}_y+\Delta \hat \sigma^i_{z}\hat \sigma^{i+2}_z.&&
	\end{eqnarray}
	Here, $\hat \sigma$ represent Pauli matrices. We choose parameters $\Delta=2$ and $\gamma=1/2$; for $\gamma={0}$ the model is integrable but still exhibits diffusive behavior \cite{Karrasch2014,DeNardis2018}.

	In Fig. \ref{fig:offset_timecorr} we show the two-time spin-spin correlations $\tr{\hat \sigma_\alpha^i(t)\sigma_\alpha^i(t')}/\mathcal D$ for $\alpha\in\{x,y,z\}$ as a function of $t$ at different $t'$, initialized in the randomly polarized Z states at $t=0$. In Fig.~\ref{fig:offset_timecorr}A we use a system size $N=16$ allowing us to benchmark CTWA with simple exact results: it is clear that dynamics are almost indistinguishable. This behavior persists at all offsets $t'$ as shown in Fig.~\ref{fig:offset_timecorr}B, and is symmetric about $|t-t'|$ as is expected. The mean-field result (colored dashed lines) does not generally reproduce the correlator, emphasizing that the initial noise is critical for the formalism. This time translation invariance is highly nontrivial, as traditional TWA methods usually break down at long times due to divergent ultra-violet noise in the system leading to spurious long time vacuum heating \cite{Blakie2008}. On the contrary, within CTWA quantum noise introduced by the initial Wigner function has a correct scaling with increasing cluster size~\cite{Wurtz2018}, and persists as a function of time: each point in phase space is generically non-stationary, as is seen in figure \ref{fig:offset_timecorr}C.2,4. This, too, is nontrivial, as initial conditions inject an amount of noise exponential in the cluster size $L$: each point is on average a distance $2^{L/2}$ from the mean. This is matched by the exponential size of the phase space $\sim4^L$. For mean-field the noise is only from thermal fluctuations; in particular it is equal to zero for each initial spin configuration (each $|\psi_n\rangle$ in Eq.~\eqref{eq:symm_corr}). In turn in generic ergodic systems such meanfield trajectories lead to relaxation of local observables to near constant (thermal) values with exponentially small fluctuations~\cite{DAlessio2015} (see Fig. \ref{fig:offset_timecorr}C.1,3).
	% If the cluster spans the whole system CTWA is exact and the time translational invariance of the system is guaranteed. For smaller cluster sizes we find that the non-linear effects do not break this time translational invariance either. In mean field dynamics the points are sampled over a very small phase space. Moreover for large cluster sizes all observables saturate to their expectation values, which almost do not change in time in ergodic systems as guaranteed by the eigenstate thermalization hypothesis~[ETH review]. And indeed as seen from Fig. XXX a) for a cluster size $L=8$ the mean field trajectory for the z-magnetization rapidly saturates to a near constant and hence the non-equal time correlation functions become nearly time-independent. This behavior should be contrasted with a typical trajectory sampled from the Gaussian ensemble (\ref{fig:offset_timecorr}B.2) which keeps fluctuating in time.
	
	We point that while the $\sigma_x \sigma_x$ and $\sigma_y \sigma_y$ time correlations decay to zero, the $\sigma_z \sigma_z$ correlation functions decay to a non-zero constant scaling as the inverse system size: $\tr{\sigma_z^i(t)\sigma_z^i(t')}/\mathcal{D}\to 1/N$ for $|t-t'|\to\infty$. This result follows from conservation of the total z magnetization: for a typical random initial state the magnetization scales as $\sqrt N$ such that the average magnetization per spin is $1/\sqrt{N}$. Within mean field different clusters cannot  exchange Z-magnetization thus the spin-spin correlation spuriously relaxes to a higher constant $1/L$ instead of $1/N$, as is seen in Fig.~\ref{fig:offset_timecorr}B.

	\begin{figure}
		\includegraphics[width=\linewidth]{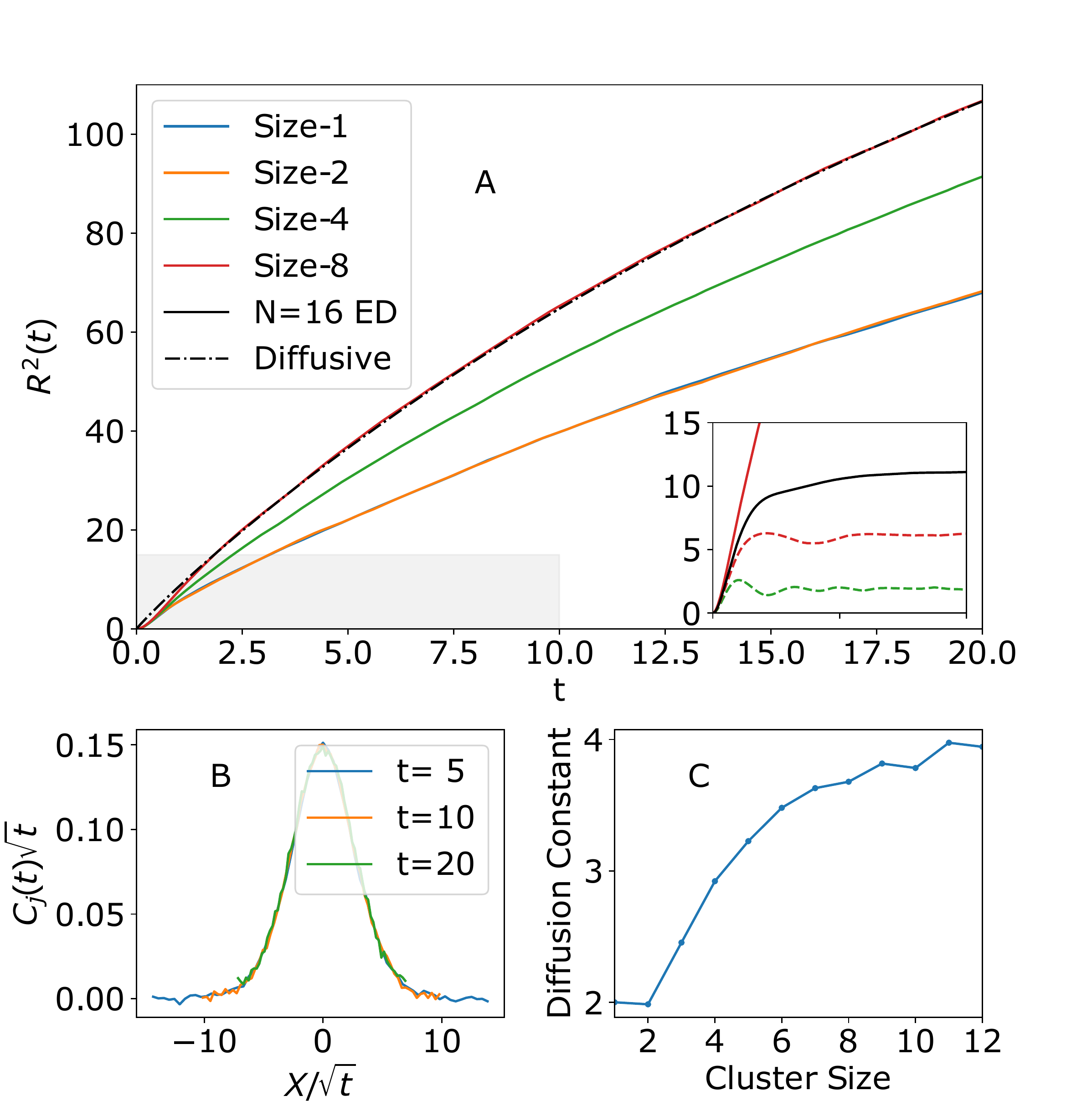}
		\caption{\textbf{Diffusive Dynamics for next-nearest neighbor XXZ chain of eq. \ref{eq:nnn_XXZ} at infinite temperature.}
			\textbf{(A)} shows the conformal width of the correlation function defined by Eq.~\eqref{eq:spreading}; Black dashed line is a single-parameter fit for classical diffusion of equation~\eqref{eq:conformal_diffusion}. Gray box and inset shows comparison of exact results for $N=16$ (solid  black line) with CTWA (red solid line) and  mean field (dashed lines) simulations for a larger system $N=64$.
			\textbf{(B)} shows scaled values of $C_{ij}$ for size-8 clusters, averaged over offsets, which takes the form of a Gaussian.
			\textbf{(C)} is a fit of the diffusion constant as a function of cluster size for $N\approx64$.}\label{fig:diffusion}
	\end{figure}
	
	Diffusion of conserved quantities at $\beta=0$ can be found using the symmetric correlator~\cite{Luitz2016a,BarLev2015}, where instead of the particle number we use the (conserved) Z-magnetization:
	\begin{equation}
	C_{ij}(t) = {1\over \mathcal D}\text{Tr}[\hat \sigma_z^i(t)\hat \sigma_z^j(0)].
	\end{equation}
	
	For diffusive systems this correlator  should be well approximated by a Gaussian whose width grows in time as $\sqrt{Dt}$, where $D$ is the diffusion constant. Therefore a natural way of extracting the diffusion constant is by computing the width of this correlation as a function of time	
	\begin{eqnarray}\label{eq:spreading}
	R^2(t)&=&\frac{\sum_{ij}{N^2\over\pi^2}\sin^2\big({\pi\over N}(i-j)\big) C_{ij}(t)}{\sum_{ij}C_{ij}(t)}
	\end{eqnarray}
	and fitting it to the solution of the classical diffusion equation (See Appendix for derivation)
	
	\begin{equation}
	\label{eq:conformal_diffusion}
	R^2(t)={N^2\over 2\pi^2}\big(1-e^{-4Dt\pi^2/N^2}\big).
	\end{equation}
	Note that in finite periodic chains systems we find it more convenient to use this conformal distance between spins; in the limit $N\to\infty$ we recover the typical Gaussian width as e.g. used in Ref.~\cite{BarLev2015}.

	In Fig.~\ref{fig:diffusion}A we show results of numerical simulations of $R^2(t)$ for different cluster sizes and the total system size $N=64$. Except for short times all the curves are well fit by the diffusion prediction~\eqref{eq:conformal_diffusion} although with a cluster dependent diffusion constant, which saturates with increasing cluster size (Fig.~\ref{fig:diffusion}C) to the asymptotic value $D\approx 3.75$. The inset shows the result of exact diagonalization for a smaller system size $N=16$ (CTWA for the same system size will be nearly identical c.f. Fig.~\ref{fig:offset_timecorr}). It is clear that the system size $N=16$ is insufficient to see diffusive behavior in this system. The panel (\ref{fig:diffusion}B) shows the correlation function $C_{ij}(t)$  rescaled by $\sqrt{t}$ with a very good collapse to the expected Gaussian profile.
	
	For size-1 clusters the Gaussian profile is expected, as the dynamics of the system is then identical to that of a classical spin chain, which is known to exhibit diffusive behavior over a wide range of parameters~\cite{Oganesyan2009a}. However, for larger cluster sizes the emergent diffusive profile is somewhat non-trivial, as the classical phase space is much larger than the naive one, encoding many ``quantum'' correlations.
	%we are dealing with a very high-dimensional classical spins with additional degrees of freedom encoding ``quantum'' correlations.
	Moreover, dependence of the diffusion constant on the cluster size $L$ (fig \ref{fig:diffusion}C) indicates that it is strongly renormalized by the underlying quantum fluctuations. As in Fig.~\ref{fig:offset_timecorr} we see that the mean field dynamics (dashed lines in the insert of fig \ref{fig:diffusion}A) is not adequate for correctly capturing long-time diffusive behavior even for relatively large cluster sizes.

	Having analyzed the diffusive spreading of correlations we now move on studying the dynamic structure factor $S(k,\omega)$ and its momentum average $S(\omega)$, containing more complete information about non-equal time spin-spin correlations: 
	\begin{eqnarray}
	S(k,\omega) &=& \sum_{ij} \int_{-\infty}^\infty dt e^{i\omega t + ik(i-j)}C_{ij}(t),\\
	S(\omega)&=&{1\over N}\sum_k S(k,\omega)= {2\pi\over N}\sum_i \int_{-\infty}^\infty dt e^{i\omega t}C_{ii}(t)\nonumber\\
	&=&{2\pi\over \mathcal D}\sum_{mm'}|\langle m'|\hat \sigma_z|m\rangle|^2\delta(\omega - E_m+E_m').\nonumber
	\end{eqnarray}
	%\blue{In Fig.~\ref{fig:Sw} we show $S(k,\omega)$ and $S(\omega)$ computed for the XXZ spin chain of size $N=64$ and different cluster sizes. {\bf JUST WANT TO MAKE SURE THAT YOU DO NOT DIVIDE BY SYSTEM SIZE AND HAVE AN EXTENSIVE $S(\omega)$} }
	
	% JW - do these two comments need to be said in their own paragraph? hmm...
	In finite size quantum systems \cite{Sachdev2011} $S(\omega)$ strictly speaking consists of isolated $\delta$-function peaks corresponding to discrete energy levels. However, as the number of states exponetially increases with the system size $S(\omega)$ effectively becomes continuous if we introduce a tiny damping factor into the time integral. We also comment that $S(k=0,\omega=0)$ diverges due to conservation of the total spin $\sigma_z$, but this divergence does not play a role at finite frequencies.
	
\begin{figure}
		\includegraphics[width=\linewidth]{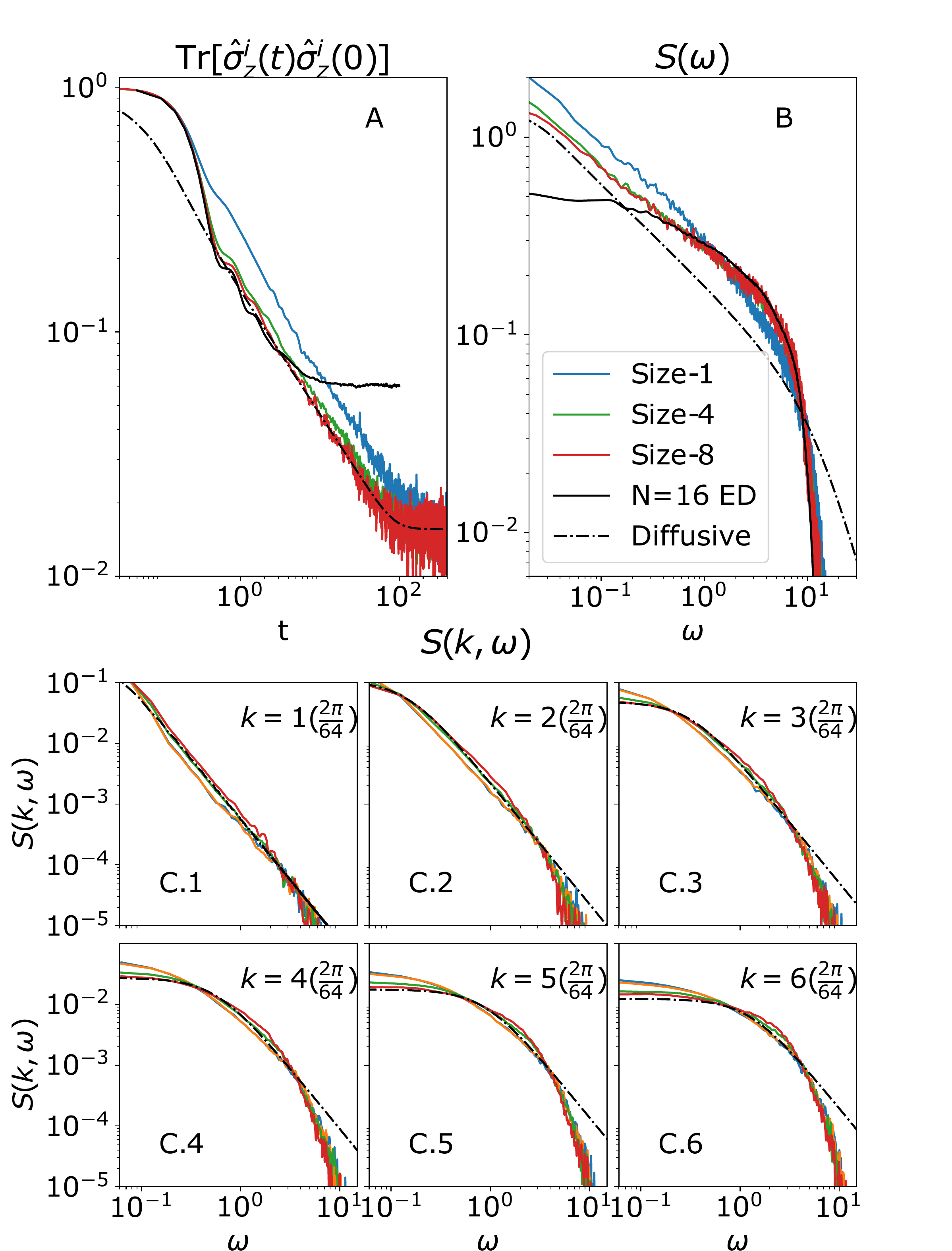}
		\caption{\textbf{Dynamic Structure Factors of the Next-Nearest Neighbor XXZ chain of eq. \ref{eq:nnn_XXZ} at infinite temperature}. \textbf{(A)} Log-log version of figure \ref{fig:offset_timecorr}A showing diffusive decay. \textbf{(B)} Momentum-averaged structure factor $S(\omega)$, which is the Fourier Transform of \textbf{(A)}. \textbf{(C.1-6)} Dynamic structure factor $S(k,\omega)$ for the first few $k$. The system size is $N=64$; dashed black lines are for classical diffusion for $D=3.75$}\label{fig:Sw}
	\end{figure}

	Figure \ref{fig:Sw}A shows the time correlations at the same site, which, after short time quantum behavior, decays diffusively as $1/\sqrt{t}$ before saturating at $1/N$. Figure \ref{fig:Sw}B shows $S(\omega)$, which is the Fourier transform of \ref{fig:Sw}A. It shows that at high frequencies the structure factor $S(\omega)$ agrees well with exact diagonalization predictions; the exponential decay as seen here is expected on general grounds \cite{DAlessio2015}. However, the simple exact diagonalization calculation fails to capture the small frequency diffusive asymptote of the structure factor $S(\omega)\propto 1/\sqrt{\omega}$~\cite{Luitz2016} due to small system sizes: there is a saturation for $S(\omega<t_c^{-1})=t_c^{1/2}$, where $t_c\sim N^2$ is the Thouless time \cite{Thouless1972}. Conversely CTWA clearly reproduces this asymptote because one can access much larger system sizes. At intermediate frequencies, there is a smooth link between the quantum and classical behaviors, allowing for a correct
	% JW - ``Correct'' seems like a very ambitious word in this context...
	behavior at all $\omega$. Figures \ref{fig:Sw}C.1-6 show the dynamic structure factor $S(k,\omega)$. for different momenta allowing one to study detailed interpolation between quantum coherent (high-frequency, short-wavelength) and hydrodynamic (low-frequency, long-wavelength) correlation functions.
	%As it is evident from the plot the exponential high frequency decay of the structure function obtained within CTWA agrees very well with the exact diaginalization prediction obtained for a smaller system sizes. This exponential decay is expected on general grounds and seen in other models (see Ref.~\cite{DAlessio2015} and Refs. therein for more details). However exact diagonalization fails to capture expected correctly small frequency diffusive asymptote of the structure factor $S(\omega)\propto 1/\sqrt{\omega}$~\cite{DAlessio2015, Luitz2016} due to small system sizes. Conversely CTWA clearly reproduces this asymptote because one can access much larger system sizes.

	\textit{Conclusion}-- We have applied CTWA to analyze infinite temperature non-equal time correlation functions in a generic XXZ chain with first and second nearest neighbor interactions. We obtained excellent agreement between the results of exact numerical simulations and CTWA predictions for small system sizes. For larger system sizes, where exact diagonalization is not available, we found that CTWA smoothly interpolates between short time quantum correlations and long time hydrodyanmic correlations. We showed that both the diffusion constant $D$ and the dynamic structure factor converge with the cluster size. Moreover as our results suggest $D$ is strongly renormalized by quantum fluctuations and can not be accurately extracted from either traditional semiclassical approaches (due to their long time failure) or exact diagonalization (due to limited system sizes). We show that mean field approaches, where one suppresses quantum fluctuations present in CTWA, gives grossly incorrect prediction for the long time behavior of the correlation functions and fails to correctly capture diffusion.	Similar incorrect hydrodynamic behavior for this model was observed in a matrix-product state based TDVP approach due to multiple conservation laws \cite{Polmann_PC2018}.
	% I am not a Latex Ninja: I can't figure out how to put this in citations. Help!
	We expect this failure  of the mean field approaches to correctly recover hydrodynamic behavior is generic and stems from relaxation of the mean field trajectories for phase space points to nearly time-independent average values. 
	
	It is interesting to see how results of our work can be extended to finite temperatures where both symmetric and anti-symmetric correlation functions are nonzero. We anticipate that at least at sufficiently high temperatures CTWA should remain accurate and allow one to extract both the dissipative and Kubo type response in strongly correlated regimes.
	
	\acknowledgements{We would like to thank E. Altman, F. Pollmann and D. Sels for stimulating discussions and additionally thank F. Pollmann for sharing unpublished results. This work was supported by NSF DMR-1813499 and AFOSR FA9550-16- 1-0334}

	\bibliographystyle{apsrev4-1}
	
	\bibliography{citationlist} 
	
	\pagebreak
		\section{Appendix: Details of CTWA}
		
		In this section we summarize key aspects of the Cluster Truncated Wigner Approximation (CTWA), which are used to obtain the results shown in the main text. For further details we refer to the paper of Ref.~\cite{Wurtz2018}.

		The CTWA is a phase space method, which approximately describes unitary dynamics in some Hilbert space via nonlinear Hamiltonian dynamics in some large-dimensional phase space. It consists of four main parts: 
		1) A definition of phase space; 2) Choice of initial conditions; 3) Proper classical Hamiltonian equations of motion defining time evolution of phase space points; and 4) Recovering information about observables and correlations. Below we briefly comment on how one implements each part.
		
		\begin{enumerate}
			
			\item	
			In CTWA phase space is associated with a set of basis operators $G=\{\hat X_\alpha\}$ which form a closed Lie algebra: $[\hat X_\alpha,\hat X_\beta]=i f_{\alpha\beta\gamma}\hat X_\gamma\in G$, where $f_{\alpha\beta\gamma}$ are the structure constants. For our system we choose the set of all operators which span clusters of spins.  For example, we can choose all independent strings of products of Pauli matrices ($\hat \sigma_x^j,\;\hat \sigma_y^j,\; \hat \sigma_z^j$) and the identity, on the sites $j$ which belong to a given cluster. For a cluster consisting of $L$ spins the total number of independent operators is $\mathcal D^2=4^L$. All traceless operators are the generators of an $SU(\mathcal D)$ group with the corresponding structure constants. Operators belonging to different clusters clearly commute with each other. Then the phase space is made by associating this set of operators to phase space variables: $\{\hat X_\alpha\}\to\{x_\alpha\}$ satisfying the canonical Poisson bracket relations defined by the same structure constants:
			\[
			\big\{x_\alpha,x_\beta\big\}=f_{\alpha\beta\gamma}x_\gamma.
			\]
			In this way all quantum operators are mapped to functions of phase space variables via Weyl quantization. In particular any operator belonging to a cluster, which can be represented through a linear combination of the basis operators maps to a corresponding linear combination of phase space point. The non-linear operators, e.g. products of basis operators belonging to different clusters map to equivalent nonlinear functions of phase space points. We note that this construction is a direct generalization of a standard quantum-classical mapping between Pauli matrices and classical spin variables. 
			
			\item
			The initial quantum state of the spins is represented by sampling an ensemble of points in phase space weighted by some probability distribution $W(x_\alpha)$, which we call the Wigner function. Although an exact Wigner Function exists, we choose a Gaussian function which reproduces the mean and variance of associated 
			basis operators. For example, a $Z$-polarized state has quantum fluctuations in $y$: $\langle(\hat \sigma^{(j)}_y)^2\rangle=1$ for any site $j$ and likewise for two sites $j,j'$ belonging to the same cluster $\langle(\hat \sigma_y^{(j)}\hat \sigma_x^{(j')})^2\rangle=1$. So when drawing initial points, the variables associated with $\hat \sigma_y^{(j)}$ and $\hat \sigma_y^{(j)}\hat \sigma_x^{(j')}$ will be drawn from a Gaussian of variance 1 and mean 0. This can be done in a general manner as detailed in Ref. \cite{Wurtz2018}. Moreover as discussed in that Ref. the actual number of independent operators scales as $2^L$, which significantly reduces complexity of sampling. 
			
			\item	Time evolution is done independently for each point in the ensemble drawn from the Wigner function. It is given by standard classical Hamiltonian equations of motion defined through the Poisson bracket:
			\[
			{\partial x_\alpha(t)\over \partial t} = f_{\alpha\beta\gamma}{\partial H(x(t))\over\partial x_\beta}x_\gamma(t)
			\]
			Within a cluster, the Hamiltonian is linear and the classical evolution gives exact quantum dynamics. Inter-cluster interactions lead to a quadratic Hamiltonian and hence to a nonlinear dynamics, which is approximate.
			As the cluster size increases, the number of nonlinear terms goes down and the CTWA dynamics becomes asymptotically exact. 
			
			\item
			Expectation values of observables and symmetric correlation functions we are interested in here are found by averaging corresponding Weyl symbols over the time-evolved phase space points (classical trajectories). In particular
			\begin{eqnarray*}
				\langle \hat A(t)\rangle &=& \overline{A(\{x(t)\})},\\
				\langle \big\{\hat A(t),\hat B(t')\big\}_+\rangle&=&2\overline{A(\{x(t)\})B(\{x(t')\})},
			\end{eqnarray*}
			where the overline denotes averaging over initial conditions drawn from the Gaussian Wigner probability distribution.	Note that because the classical equations of motion are generally nonlinear averaging over the initial conditions and time propagation are noncommuting operations.

		\end{enumerate}

		\section{Appendix: Exactness of the single cluster CTWA }
		
		In Ref.~\cite{Wurtz2018} we mentioned that CTWA exactly reproduces not only expectation values of observables but also their non-equal time correlation functions in the limit when the cluster size becomes equal to the system size and hence the evolution becomes linear. Let us provide here a simple proof of this statement for the symmetric correlation functions. In the linear case time evolution of arbitrary phase space point operator is simply a unitary rotation given by some generally time dependent unitary matrix $U_{\alpha\beta}(t)$ and therefore both quantum operators and classical phase space points evolve in the same way
		\[
		\hat X_\alpha(t)=\sum_\beta U_{\alpha\beta}(t) \hat X_\beta(0),\quad x_\alpha(t)=\sum_\beta U_{\alpha\beta}(t) \hat x_\beta(0)
		\]
		For a time independent Hamiltonian the unitary is given by the exponent of the effective magnetic torque:
		\be
		U_{\alpha\beta}(t)=\mathrm e^{M_{\alpha\beta} t},\quad M_{\alpha\beta}=f_{\alpha\beta\gamma} B_\gamma,
		\ee
		where the magnetic field $\vec B$ is defined as usual according to $\hat H=-\sum_\alpha B_\alpha \hat X_\alpha$. For a time-dependent Hamiltonian the unitary $U_{\alpha\beta}$ still exists but is defined through a more complicated time-ordered exponential of the time integral of the magnetic field. From this we find	
		\begin{multline}
		\langle \{\hat X_\alpha(t),X_\beta (t')\}_+\rangle = \sum_{\gamma,\delta} U_{\alpha\gamma}(t) U_{\beta\delta}(t') \langle \{\hat X_\gamma(0),X_\delta (0)\}\rangle\\
		= \sum_{\gamma,\delta} U_{\alpha\gamma}(t) U_{\beta\delta}(t') \overline{2 x_\gamma(0) x_\delta(0)}=\overline{2 x_\alpha(t)x_\beta(t')}.
		\end{multline}
		Note that even fat the level of a single cluster fluctuations in initial conditions are crucial for correctly reproducing the non-equal time correlation functions. On the contrary, mean field approximation would generally fail to predict such correlation functions.

		\section{Appendix: Solution of the Diffusion Equation on a Discrete Lattice}
		
		In this appendix we detail derivation of Eq. \ref{eq:conformal_diffusion}, as well as of the expressions representing the black dashed line of figures \ref{fig:diffusion} and \ref{fig:Sw}. The discrete classical diffusion equation reads:
		\begin{equation}
		\partial_t\rho_i = -D(2\rho_l - \rho_{l-1} - \rho_{l+1}).
		\end{equation}
		Here $\rho$ represents a conserved change, which is given by the Z-magnetization in our case. This eqution can be easily solved in the momentum space using the Fourier transform of $\rho$:
		\[
		\eta_k = \sum_l e^{ikl}\rho_l, \quad k=0,2\pi/N,\dots 2\pi(N-1)/N.
		\]
		Then the diffusion equation for each Fourier component $\eta_k$ reduces to a simple first order differential equations, which is easy to solve:
		\begin{multline}
		\partial_t \eta_k=-2D\rho_k(1-\cos(k)),\\
		\Rightarrow\;
		\eta_k(t) = \eta_k(0)e^{-2Dt(1-\cos(k))}.
		\end{multline}
		
		Using this solution one can easily find the conformal diffusion width shown in the main text (Eq. \ref{eq:spreading}):
		\begin{multline}
		R^2(t)=\sum_{l}{N^2\over\pi^2}\sin^2\bigg({\pi j\over N}\bigg)\rho_j(t)\\
		=\sum_{kl}{N^2\over\pi^2}\sin^2\bigg({\pi l\over N}\bigg)e^{-ikl}\eta_k(t)\\
		={N\over 2\pi^2}\bigg(\eta_0 - {1\over 2}\eta_{2\pi/N} - {1\over 2}\eta_{-2\pi/N}\bigg).\nonumber
		\end{multline}
		Inserting the explicit solution for form of $\eta_k(t)$ with the initial condition  $\eta_k(0) = 1$ and expanding $\cos(2\pi/N)\approx 1-(2\pi/N)^2/2$ at large $N$ we derive Eq. \ref{eq:conformal_diffusion} from the main text. Similarly, one can find the diffusive structure factor	$S(k,\omega)$:
		
		\begin{equation}
		S(k,\omega) =\int_{-\infty}^{\infty}dt e^{i\omega t}\eta_k(t)
		= \frac{4 D(1-\cos(k))}{\omega^2 + 4 D^2 (1-\cos(k))^2}.
		\end{equation}

\end{document}